\documentclass[12pt,preprint]{emulateapj}
\usepackage{graphicx}
\usepackage{natbib}
\usepackage{amsmath}
\usepackage{placeins}
\usepackage{color}

\newcommand{\degree}{\ensuremath{^\circ}\,}

\def\MJ{{M$_{Jup}$\,}}
\def\MS{{M$_{\odot}$\,}}

\def\gtaprx{ \mathrel{ \vcenter{
      \offinterlineskip \hbox{$>$}
      \kern 0.3ex \hbox{$\sim$}    } } }

\def\ltaprx{ \mathrel{ \vcenter{
      \offinterlineskip \hbox{$<$}
      \kern 0.3ex \hbox{$\sim$}    } } }
      
\def\apj{{ApJ}}

\def\aap{{A\&A}}
\def\pasp{{PASP}}

\def\mnras{{MNRAS}}
\def\nat{{Nature}}

\def\icarus{{Icarus}}

\shorttitle{Habitability of Exomoons at the Hill or Tidal Locking Radius}
\shortauthors{Natalie R. Hinkel \& Stephen R. Kane}
\slugcomment{Submitted for publication in the Astrophysical Journal}

\begin{document}

\title{Habitability of Exomoons at the Hill or Tidal Locking Radius}
\author{Natalie R. Hinkel \& Stephen R. Kane}
\affil{NASA Exoplanet Science Institute, Caltech, MS 100-22, 770
  South Wilson Avenue, Pasadena, CA 91125, USA}
\email{natalie.hinkel@gmail.com}

\begin{abstract}

Moons orbiting extrasolar planets are the next class of object to be observed and characterized for possible habitability.  Like the host-planets to their host-star, exomoons have a limiting radius at which they may be gravitationally bound, or the Hill radius.  In addition, they also have a distance at which they will become tidally locked and therefore in synchronous rotation with the planet.  We have examined the flux phase profile of a simulated, hypothetical moon orbiting at a distant radius around the confirmed exoplanets $\mu$ Ara b, HD 28185 b, BD +14 4559 b, and HD 73534 b.  The irradiated flux on a moon at it's furthest, stable distance from the planet achieves it's largest flux gradient, which places a limit on the flux ranges expected for subsequent (observed) moons closer in orbit to the planet.  We have also analyzed the effect of planetary eccentricity on the flux on the moon, examining planets that traverse the habitable zone either fully or partially during their orbit.  Looking solely at the stellar contributions, we find that moons around planets that are totally within the habitable zone experience thermal equilibrium temperatures above the runaway greenhouse limit, requiring a small heat redistribution efficiency.  In contrast, exomoons orbiting planets that only spend a fraction of their time within the habitable zone require a heat redistribution efficiency near 100\% in order to achieve  temperatures suitable for habitability. 
  Meaning, a planet does not need to spend its entire orbit within the habitable zone in order for the exomoon to be habitable.  Because the applied systems are comprised of giant planets around bright stars, we believe that the transit detection method is most likely to yield an exomoon discovery.

\end{abstract}

\keywords{stars: planetary systems, planets and satellites: individual ($\mu$ Ara b, HD 28185 b, BD +14 4559 b, and HD 73534 b), astrobiology -- planetary systems}

\section{Introduction}
\label{intro}

As the search to understand and characterize exoplanets expands, so too will the detection limits of the telescopes employed to observe the distant systems.  It therefore seems logical to expect that scientists in the near future will be observing exomoons.  While there has yet to be an exomoon detection, there has already been some research into the analysis needed for a systematic exomoon search \citep{Kipping2012}.  The possible formation scenarios and frequency predictions by \citet{Morishima2010} and \citet[][and references therein]{Elser2011} are indicative of the number of exomoons expected in our local neighborhood, not to mention the 166 satellites orbiting the 8 planets within the Solar System, exomoons are the next environment to search for signs of habitability.

There are two major gravitational constraints on exomoons
beyond the Roche limit:
one that binds the moon to the host-planet and one that induces synchronous rotation with the planet.  The distance to which these limits extend are measured by the Hill radius 
(or fraction thereof depending on orbital stability, see \S \ref{radii})
and the tidal locking radius, respectively.  By studying moons at these more extreme distances, we are able to better understand the influence of the host-star as well as the host-planet on the flux variations on the moon.  Exomoons at large distances reach both the 
closest and furthest points from the host-star and achieve the greatest possible flux gradients during the course of one planetary phase.  In addition, if the host-planet traverses outside of the habitable zone for a portion of its orbit, we may also analyze those distance effects with respect to the flux experienced on the moon.  
While the larger distances may diminish the overall effects of the planet's contribution to the irradiation felt on the moon, we have chosen to focus our study on the analysis of the moon-planet-star orbit geometry and how it affects the flux received on the moon.  In this way, we are able to estimate whether a moon may be habitable within the given system.
Understanding the 
more extreme conditions

of these extended systems places a limit on the habitability conditions of yet undiscovered moons likely located at shorter distances from both the host-star and host-planet.  

The purpose of this paper is to examine the effect of a large planet-moon distance (Hill or tidal locking radius) on the flux on the moon's surface and analyze whether a moon might still be habitable when the host-planet is not fully within the habitable zone.  In \S \ref{temp} we discuss the influences on the moon's flux phase profile, such as illumination and radiation, planet eccentricity, and distance from the host-planet.  In \S \ref{apps}, we examine the flux profile for a hypothetical moon at either the Hill or tidal locking radius orbiting $\mu$ Ara b, HD 28185 b, BD +14 4559 b, and HD 73534 b.  
We have chosen these systems specifically to explore a variety of planet masses, such that the planet could host larger moons assuming that the scale of the moon scales with the mass of the primary planet.  We have also analyzed variations in eccentricities and time spent within the habitable zone.
In \S \ref{hab} we discuss the effect on the exomoon habitability by each of the four application systems 
as well as potential extreme thermal contributions from the host planet on the exomoon. 
In \S \ref{detect} we discuss the possibility of future detections for exomoons using modern techniques.  Finally, we summarize our results in \S \ref{sum}.

\section{Exomoon Flux Dependencies}
\label{temp}
One of the major components to habitability on an exomoon, similar to that on a planet, is the flux range experienced by the moon.  The flux is influenced by the luminosity of the host-star, the distance of the planet-moon system to the star, and the distance of the moon to the planet.  We discuss how these major contributors effect the moon flux below.

\subsection{Illumination and Radiation}
\label{illrad}
In order to take into account multiple sources of radiation, \citet{Heller2013} developed a model that determines the effect of stellar illumination, light from the star reflected off of the planet, the planet's thermal radiation, and tidal heating on the surface of an exomoon.  Their code allowed for the specification of a variety of system parameters, such as stellar host-mass (M$_s$), planet mass (M$_p$), moon mass (M$_m$), star radius (R$_s$), planet radius (R$_p$), stellar effective temperature (T$_{eff}$), bond albedo of the planet ($\alpha_p$), rock-to-mass fraction of the moon ($rmf$), semi-major axis of the planet to the star ($a_{sp}$), eccentricity of the planet (e$_{sp}$), semi-major axis of the moon about the planet ($a_{pm}$), eccentricity of the moon's orbit ($e_{pm}$), inclination of the moon with respect to the planet ($i$), and orientation of the inclined orbit with respect to the periastron ($\omega$).  With these prescriptions, they were able to calculate the total flux received on the moon during one moon orbital phase (or ``phase curves"), the average flux with respect to the moon's latitude and longitude after one complete planet revolution (or ``flux map"), as well as the physical orbit of the planet and moon as they orbit the host-star (or ``orbital path").  

While the \citet{Heller2013} code was very extensive and thorough, there are some limitations to the possible applications.  For example, the authors assumed that the heat from gravitationally-induced shrinking of the gaseous planet was negligible compared to the stellar luminosity.  In a similar vein, the ratio of the period of the moon about the planet ($P_{pm}$) to the period of the planet about the star ($P_{sp}$) follows, $P_{pm} / P_{sp} \ltaprx$ 1/9, as per \citet{Kipping2009} -- see \S \ref{radii} for more discussion.  Also, the moon is assumed to be in a relatively circular orbit due to the effects of tidal heating dampening the orbital eccentricities, unless the moon is perturbed by the interactions with other bodies.  In their paper, \citet{Heller2013} explored $e_{pm} \le 0.05$.  Finally, all moons are assumed to be tidally locked, the moon never experiences a penumbra from the planet, $M_{p} >>  M_{m}$, and both atmospheric and geologic contributions are ignored.  We suggest the reader confers with the original paper for a more thorough description of the techniques and simplifications employed by \citet{Heller2013}.

\begin{figure}
  \includegraphics[width=9.3cm]{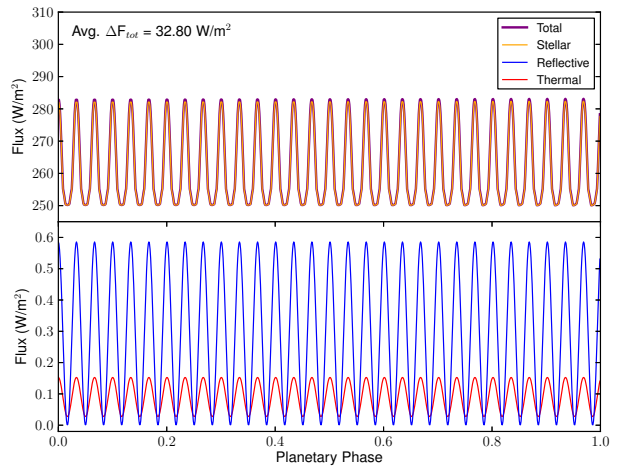}
  \caption{Flux phase profile of an example Earth-sized moon orbiting a Jupiter-sized planet, at 1\,AU from the Sun, for one year.  Because of the large range covered by the thermal, reflective, and stellar fluxes, we have broken the plot into two range regimes, to emphasize the contribution of all parts to the total flux.  The planet's orbit has an eccentricity of 0.0.  The inclination of the moon to the planet is $i = 0 \degree$ and the planet-moon semi-major axis is 0.01 AU.  The average total flux variation on the moon is given in the top left corner.}
  \label{ex1}
\end{figure}

Using their publicly available code (see their Appendix C), we were able to adjust their methodology to calculate the average flux experienced on the moon's surface (across all latitudes and longitudes) over the course of one planet revolution.  The flux experienced on the moon was calculated assuming blackbody radiation emitted by the star.  In Fig. \ref{ex1} we show a year-averaged flux profile for an example system featuring an Earth-sized moon orbiting a Jupiter-sized planet, at 1\,AU from the Sun.  The different color lines show heat contributions from the star (orange), planetary reflection (blue), and planetary thermal emission (red) which are summed to the total (purple).  
We have split the figure into two flux-range regimes in order to better illustrate the contribution from all three sources, while noting that all three components (stellar, reflective, and thermal) lead to the total.
The average total flux oscillation experienced on the moon as a result of its orbit around the planet is given in the top left corner of the plot.  Details of this example system are found in Table \ref{tab.params}, such that the eccentricity of the planet e$_{sp} = 0.0$.  The inclination of the moon is $i = 0 \degree$ and the planet-moon semi-major axis is 0.01 AU.  The moon's orbit has an eccentricity of e$_{pm} = 0.0$, meaning that tidal heating has no effect on the flux of the moon.  The total flux of the moon in Fig. \ref{ex1} is dominated by the stellar luminosity, with smaller oscillations due to the revolution of the moon about the planet, as can be seen in the reflected light off the planet.  

\subsection{High Eccentricity}
\begin{figure*}
  \begin{center}
  \includegraphics[width=18cm]{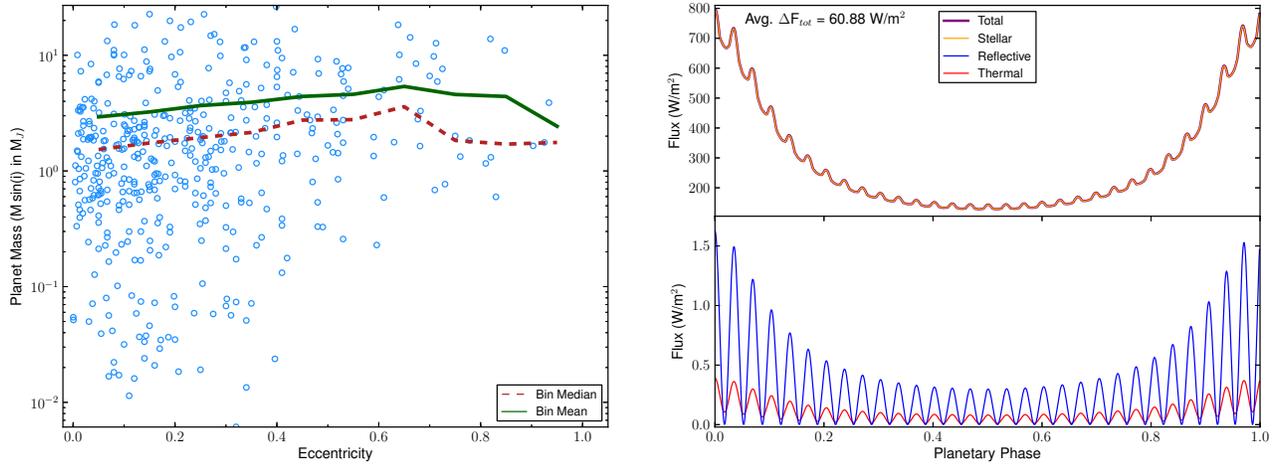}
  \end{center}
  \caption{On the left, the mass vs. eccentricity distribution of all confirmed exoplanets.  The median (red) and mean (green) of the masses in eccentricity bins of 0.1 are overlaid to show the general increase of mass with eccentricity.  The figure on the right is similar to Fig. \ref{ex1}, where the planet's eccentricity is now $e = 0.4$.}
  \label{ex2}
\end{figure*}

As the number of confirmed exoplanets increases, so does the ability to characterize the systems and determine widespread patterns.  The eccentricity of the host-planet has a direct consequence on the flux on the exomoon, since it affects the distance between the host-star and the planet-moon system, as studied by \citet{Heller12}.  Figure \ref{ex2} (left) shows all of the confirmed RV exoplanets with measurements for both planet mass and eccentricity, totaling 441 planets.  Overlaid on the plot are the median (red) and mean (green) masses as determined for 0.1 eccentricity bins.  The median within a bin rules out outlier masses and the mean determines the average per bin.  Both trends indicate that as eccentricity increases until $e_{sp} \sim $ 0.7, planet mass increases.  For $e_{sp} > $0.7, planet mass tends to decrease.
The RV-technique is biased toward the lower-eccentricity planets, such that the number of planets decreases with higher eccentricities.  In addition, there is also a bias in observing higher mass planets.  Despite these idiosyncrasies in the data, we wish to study what is currently considered a ``typical" system in the high-eccentricity regime, bearing in mind that this definition may change in the future.

We continue with the example system in Fig. \ref{ex2} (right), where we have changed the eccentricity, e$_{sp} = 0.4$, which is on par for a more massive planet.  The high eccentricity dominates the moon's flux phase profile such that it is clear when the planet-moon system is at periastron (planetary phase = 0.0) and when it is at apastron (planetary phase = 0.5).  Due to the close proximity of the planet-moon to the host-star, not only does the moon's flux start +500 W/m$^2$ higher than at low eccentricity (see Fig. \ref{ex1}), but the average total small-scale flux variation is higher by 28.08 W/m$^2$.  On the other hand, the average flux on the moon at low eccentricity is 262 W/m$^2$ while at high eccentricity it is 283 W/m$^2$, so not significantly different.

\subsection{Hill radius vs. Tidal Locking radius}
\label{radii}
The final system variable that has a large impact on the flux on the moon's surface is the distance between the planet and the moon, where the semi-major axis is $a_{pm}$.  As we have seen in the previous two incarnations of our example system, a moon that is relatively close to the planet experiences somewhat small flux fluctuations as a result of its close orbit.  Since there have not been any observations of exomoons to-date, we wish to explore the limits of gravitational influence on the moon.

We first consider the radius at which a moon is gravitationally bound to a planet, due to the influence of the nearby host-star, or the Hill radius:
\begin{equation}\label{hill}
r_H = a_{sp}\, \chi \,(1-e_{sp})\, \sqrt[3]{\frac{M_p}{3\,M_s}} .
\end{equation}
Here $\chi$ is an observational factor implemented to take into account the fact that the Hill radius is just an estimate and that other effects may impact the gravitational stability of the system.  Following both \citet{Barnes2002} and \citet{Kipping2009}, a conservative estimate is that $\chi \ltaprx 1/3$.  We choose to use $\chi \sim 1/3$ such that $P_{pm} / P_{sp} \ltaprx$ 1/9 \citep{Domingos06, Kipping2009}.  Meaning, a moon beyond this cautious radius would be perturbed by tidal interaction with the star and would not be able to maintain a stable orbit around the planet.  

Second, we look at the distance at which a planet's tidal influence induces the moon to become locked in a synchronous rotation.  The tidal locking radius is defined as:
\begin{equation}\label{tidal}
r_{TL} \approx \bigg(\frac{3\, G \,M^2_p\, k_2\, R^5_m \,t_L}{\omega_{0}\, I \, Q}\bigg)^{1/6}, 
\end{equation}
where $G$ is the gravitational constant, $k_2 = 0.3$ is the second-order Love number, $t_L$ is the timescale for the moon to become locked, $\omega_{0}$ is the initial spin rate of the moon in radians per second, $I \approx 0.4\,M_m\,R^2_m$ is the moon's moment of inertia, and $Q$ is the dissipation function of the moon \citep{Peale1977, Gladman1996}.  From \citet{Kasting:1993p6301}, we take $\omega_{0} =$ one rotation per 13.5 hrs and Q = 100 for a conservative estimate.  Based on the present age of the Earth, $t_L = $ 4.5 Gyr.

\begin{figure*}
  \begin{center}
  \includegraphics[width=18cm]{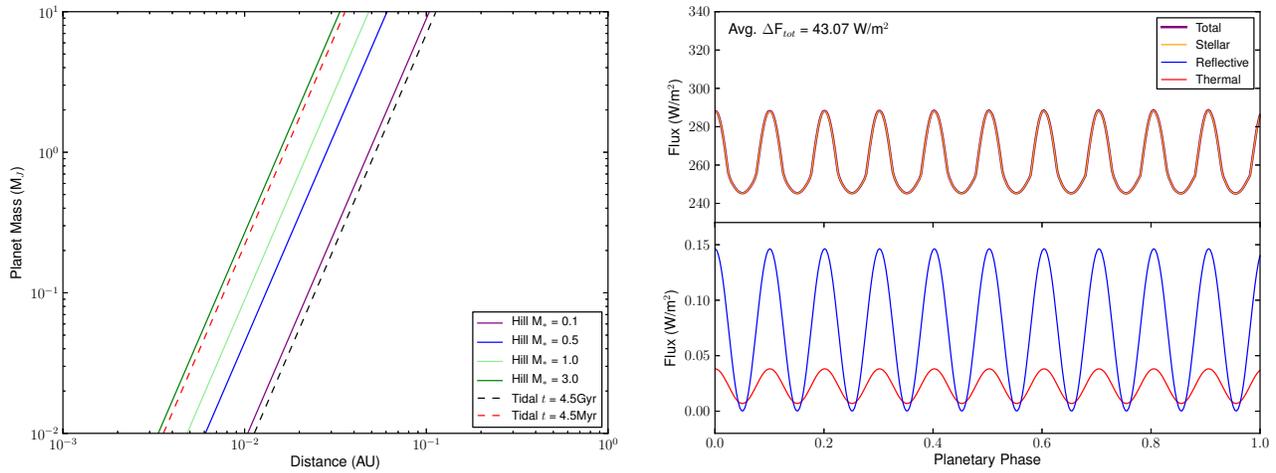}
  \end{center}
  \caption{On the left, planet mass at 1 AU as it pertains to the Hill radii with respect to different solar masses (solid lines) and tidal locking radii based on different timescales (dashed lines).  To the right, similar to Fig. \ref{ex1} where the planet-moon semi-major axis is equivalent to the Hill radius for a 1M$_\odot$ star, or a = 0.02 AU.}
  \label{ex3}
\end{figure*}

In Fig. \ref{ex3} (left) we analyze the Hill and tidal locking radii with respect to planet mass in a log-log plot.  The solid lines show the Hill radii for a variety of stellar masses, while the dashed lines give the tidal locking radius when $t_L$ = 4.5 Gyr (black) and 4.5 Myr (red).  As planetary mass increases, so does the distance at which gravity keeps both the moon bound and tidally locked.  We will be using the typically accepted value $t_L = $ 4.5 Gyr, but find it interesting to consider the range of values produced by altering this variable, especially with respect to the Hill radius, for different stellar masses.  For all solar and planet masses at a distance of 1 AU, particularly the 1.0 M$_{\odot}$ star and 1.0 M$_{Jup}$ planet in our example system, the Hill radius is smaller than the tidal locking radius when $t_L$ = 4.5 Gyr.

Revisiting our example system, we have plotted the flux profile with respect to planetary phase in Fig. \ref{ex3} (right).  However, in this instance we have placed the planet near the Hill radius, $r_H =$ 0.022 AU, where $a_{pm} =$ 0.02 AU.  For this system, the orbit of the moon dominates the flux profile such that $P_{pm} / P_{sp}$ = 1/10, which is within the bound determined by \citet{Kipping2009} and \citet{Heller2013}.  The average total flux fluctuation, $\sim$ 43.07 W/m$^2$, is smaller in this example system compared to the same system at various planetary eccentricities, most likely as a result of the larger $a_{pm}$.  The average total flux is also similar to the previous incarnations of the example system at 262 W/m$^2$.  

We have analyzed the hypothetical exomoon in three scenarios that only vary slightly from one another.  When the eccentricity of the planet-moon system is large, the surface of the exomoon experiences much more extreme flux variations compared to a circular orbit.  The average flux of the exomoon at a large eccentricity is also slightly larger at high eccentricity compared to $e$ = 0.  Comparatively, when the exomoon is placed at a distance further from the planet in accordance with the Hill (in this case) or tidal radius, the average flux on the surface is the same as the original scenario.  In other words, the thermal and reflected radiation from the planet do not make a significant contribution to the flux range experienced on the moon, while the eccentricity of the planet-moon system results in an extreme range of fluxes and slightly higher average flux.

\section{Applications to Known Systems}
\label{apps}
\begin{figure}
  \begin{center}
  \includegraphics[width=5.5cm]{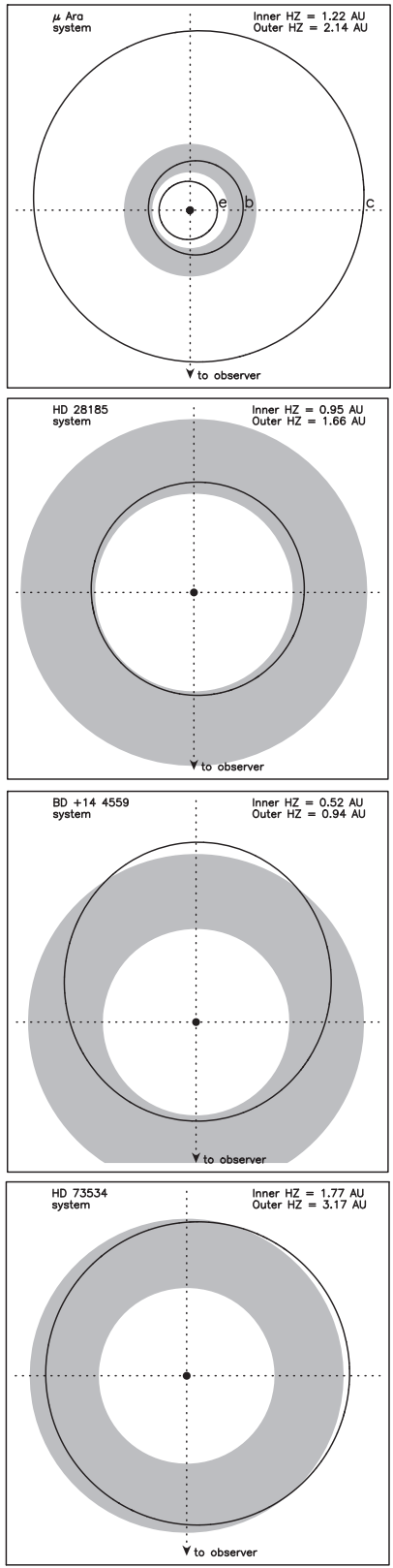}
  \end{center}
  \caption{The planetary orbits (multi planets labeled for $mu$ Ara, where the inner d-planet is extremely close to the host-star) and habitable zone regions (shaded gray).  The inner and outer HZ radii are given in the top-right corner.}
  \label{hzs}
\end{figure}

Using the techniques described in \S \ref{temp}, we analyze the flux profiles for hypothetical moons in known exoplanetary systems.  In order to better compare the applications to each other, we maintain a similar planet-moon binary, where R$_p$ = 1.0 R$_{Jup}$, $\alpha_p$ = 0.343, M$_m$ = 10 $\times$ the mass of Ganymede = 0.25 M$_{\oplus}$, R$_m$ = 0.68 R$_{\oplus}$, moon $rmf$ = 0.68, e$_{pm}$ = 0.0, $i$ = 0\degree, and $\omega$ = 0\degree.  We have set the albedo of the planet and satellite equal to each other, similar to \citet{Heller2013}.  Only a$_{pm}$ is varied between the systems per the Hill or tidal locking radius.  The parameters for each application can be found in Table \ref{tab.params}, while a pictorial view of each system are shown in Fig. \ref{hzs} with the calculated habitable zones in gray \citep{Kopparapu2013}.  
In addition, we have plotted multiple stellar masses in the left-hand plots of Figs. \ref{muAra} -- \ref{hd7} (despite the stellar masses within each system being well understood) in order to demonstrate how the Hill and tidal locking radii change given the orbital parameters of a system.  By plotting all of the applications on the same scale, it is easier to understand the interplay between the two extreme radii for the various systems, especially in the case of HD 73534 b.

\subsection{$\mu$ Ara b}
The $\mu$ Ara system contains four confirmed exoplanets around a 1.15 \MS G3IV-V type-star \citep{Pepe2007}, see Fig. \ref{hzs} (top) where the d-planet is too close to the host-star to be distinguished.  While the c-planet is the most massive at 1.89 \MJ, it is also the furthest away such that a$_{sp}$ = 5.34 AU.  The 1.746 \MJ b-planet is the only planet to be within the habitable zone, or between [1.22, 2.14] AU, for the entirety of its orbit, where a$_{sp}$ = 1.527 AU and e$_{sp}$ = 0.128.
We chose to study the $\mu$ Ara system because of the presence of numerous, relatively massive, planets.  The large mass of the b-planet results in a large Hill sphere which presents many possibilities for a moon system, both in terms of masses and moon-planet separations.  In addition, the eccentricity of the b-planet is enough such that it may vary the equilibrium temperature of the planet-moon system by a substantial amount, while remaining fully within the habitable zone.

\begin{figure*}
  \begin{center}
  \includegraphics[width=18cm]{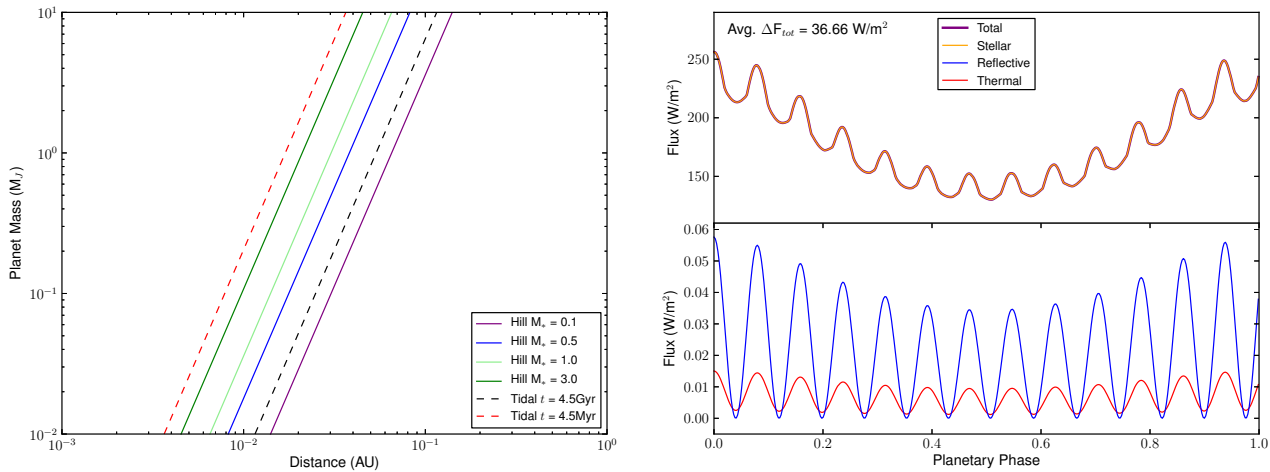}
  \end{center}
  \caption{Similar to Fig. \ref{ex3} (left) and Fig. \ref{ex1}, respectively, but for the $\mu$ Ara b system.  On the right, a Hill radius of a = 0.03 AU has been applied as the semi-major axis to a hypothetical moon around $\mu$ Ara b with mass 10 $\times$ the mass of Ganymede = 0.25 M$_{\oplus}$, R$_m$ = 0.68 R$_{\oplus}$, $i = 0 \degree$, and Jupiter albedo.}
  \label{muAra}
\end{figure*}

In Fig. \ref{muAra} (left) we explore the maximum distance that a hypothetical 0.25 M$_{\oplus}$ moon could orbit stably and remain tidally locked.  From the plot, the tidal locking radius is more dominant in this system, making the Hill radius for a near solar-mass star the limiting distance for the placement of the hypothetical moon.  The Hill radius for $\mu$ Ara b is $r_H = 0.035$ AU, therefore, to ensure the planet is fully within the Hill sphere, a$_{pm}$ = 0.03 AU.

The flux profile for the hypothetical moon near the Hill radius is shown in Fig. \ref{muAra} (right).  Similar to Fig. \ref{ex3}, the large orbital period of the moon has a strong influence on the flux phase fluctuations.  However, the non-zero eccentricity of the planet-moon system is also apparent in the flux profile, much like Fig. \ref{ex2} but not as exaggerated.  The moon's surface experiences a minimum flux of 130 W/m$^2$ and a maximum of 256 W/m$^2$, with the average total flux oscillating $\sim$37 W/m$^2$ as the moon's phase changes.

\subsection{HD 28185 b}
While the HD 28185 system only contains one confirmed exoplanet  \citep{Minniti2009}, that planet is 5.8 \MJ \,and maintains an orbit that is fully within the habitable zone at a$_{sp}$ = 1.023 AU and e$_{sp}$ = 0.05 (Table \ref{tab.params}). 
Given the mass extent of the confirmed exoplanet, we wanted to explore the theoretical moon-planet interaction within the enormous Hill sphere.  However, per one of the four simplifications noted in \citet{Heller2013}, we had to ensure that the distance between the moon and the planet was not so large that P$_{pm} \approx$ P$_{sp}$.  Therefore, a system with a large planet had to be found that also fell nearer the inner habitable zone boundary.  In contrast to $\mu$ Ara, HD 28185 b has a very low eccentricity, such that the flux received on the moon will be affected the most by the distance of the planet from the moon.
The habitable zone around the host-star, a G5V type-star with a mass of 0.990 \MS, extends from 0.95 -- 1.66 AU (Fig. \ref{hzs}, second from the top).

\begin{figure*}
  \begin{center}
  \includegraphics[width=18cm]{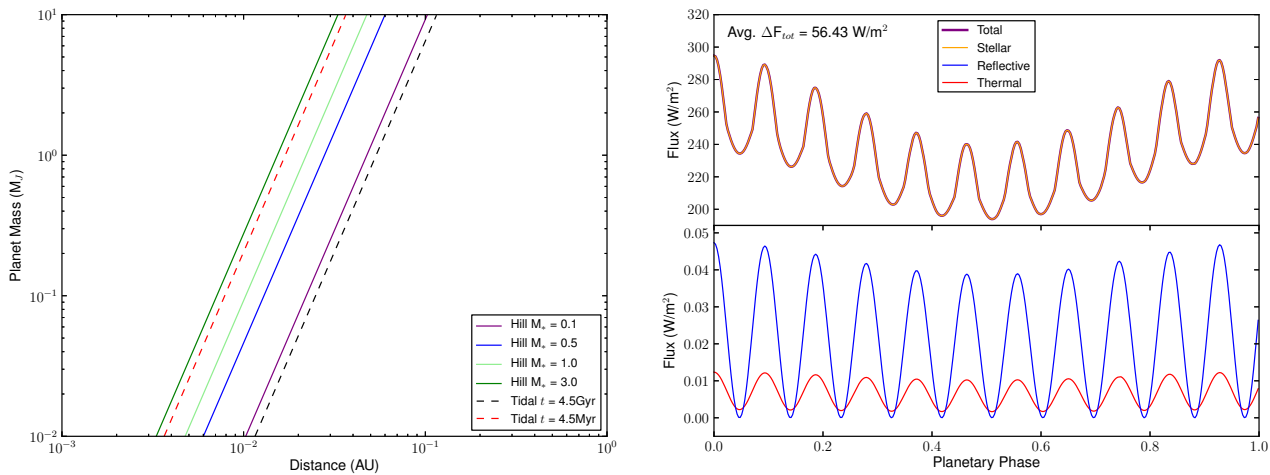}
  \end{center}
  \caption{Similar to Fig. \ref{muAra} but for HD 28185 b, where the hypothetical moon is near the Hill radius at a = 0.035 AU.}
  \label{hd2} 
\end{figure*}

Figure \ref{hd2} (left) is similar to Fig. \ref{muAra} (left) in that it examines the distance limit for a hypothetical moon as defined by either the Hill radius or the tidal locking radius for the system.  Like $\mu$ Ara, the gravitational boundary for HD 28185 b is more restricted by the Hill radius for a 1.0 \MS star as compared to the tidal locking radius at $t_L$ = 4.5 Gyr.  The Hill radius is r$_H$ = 0.039 AU, where we define a$_{pm}$ = 0.035 AU in order to guarantee the moon is fully within the Hill radius.  

In Fig. \ref{hd2} (right) we show the flux profile for the hypothetical moon around HD 28185 b.  Because of the relatively low planetary eccentricity, the moon's orbit dominates the flux fluctuations seen on the surface.  In this scenario, $P_{pm} / P_{sp} \sim$ 1/11.  The exoplanet orbiting HD 28185 is one of the largest planets that spends the entirety of its orbit within the habitable zone.  Yet, despite it's large size, the thermal and reflected flux contributions, $\sim$0.01 W/m$^2$ and $\sim$0.05 W/m$^2$, respectively, from this planet is nearly identical to that from $\mu$ Ara b.  Therefore, the larger average flux variations seen on the moon's surface, $\sim$56 W/m$^2$, are most likely due to the larger semi-major axis between the moon and the planet.

\renewcommand*{\thefootnote}{\fnsymbol{footnote}}

\subsection{BD +14 4559 b}
Unlike the previous applications, we wanted to analyze a system that did not spend the entirety of its period within the habitable zone.  We chose to analyze the BD +14 4559 system because the planet has a highly eccentric orbit, e$_{sp}$ = 0.29 and only spends 68.5\% of its orbital phase within the habitable zone (Fig. \ref{hzs}, second from the bottom).  The single confirmed exoplanet orbiting BD +14 4559 has a mass that is 1.52 \MJ \citep{Niedzielski2009}, which places it between both the median and mean lines in Fig. \ref{ex2} (left).  Taking into account the observational bias in the RV technique which preferentially detects planets with small eccentricity, we expect this to be a standard eccentricity for an RV-observed planet of this mass.  In this way, we could explore the flux received on the exomoon in a ``standard" high-eccentricity system, despite the host planet being far from the host star for a large fraction of its orbit.
The observed b-planet has an orbit at 0.776 AU from the 0.86 \MS K2V-type star (Table \ref{tab.params}), where the boundaries of the habitable zone fall between 0.52 and 0.94 AU.

\begin{figure*}
  \begin{center}
  \includegraphics[width=18cm]{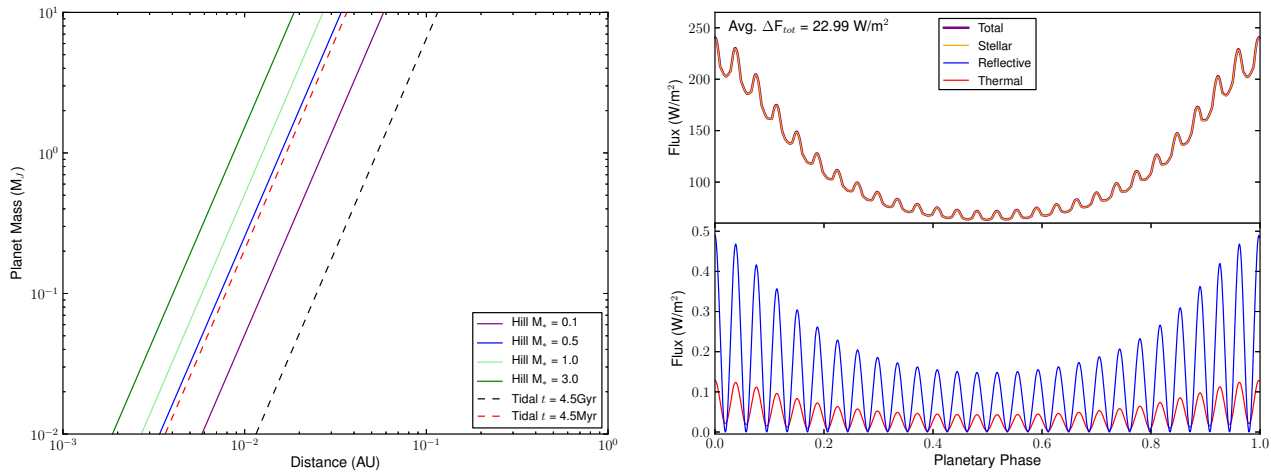}
  \end{center}
  \caption{Similar to Fig. \ref{muAra} but for BD +14 1559 b, which spends $\sim$68.5\% of it's orbital phase in the habitable zone \citep{Kane2012}, with the hypothetical moon is near the Hill radius at a = 0.01 AU.}
  \label{bd4}
\end{figure*}

Due to the higher eccentricity, the distance to the Hill radius is smaller than those seen in the other applications (Fig. \ref{bd4}, left).  In contrast, the tidal locking radius has not shifted significantly, making the two radii even more discrepant.  Similar to both $\mu$ Ara b and HD 28185 b, the hypothetical moon orbiting the b-planet is most limited by the Hill radius, r$_H$ = 0.015 AU.  For our purposes, we place the hypothetical 0.25 M$_{\oplus}$ moon at a distance of a$_{pm}$ = 0.01 AU.

The closer proximity of the hypothetical moon to the planet means that $P_{pm} / P_{sp} \sim$ 1/27, as shown in Fig. \ref{bd4} (right).  The shorter moon orbital period is most likely the cause of the smaller average total flux fluctuation, $\sim$23 W/m$^2$.  As a result, the relatively large planetary eccentricity dominates the flux profile, which swings from 241 W/m$^2$ to 63 W/m$^2$.

\subsection{HD 73534 b}
Out of the +440 confirmed RV planets with planet mass and eccentricity measurements, only 35 of them had a Hill radius at a further distance than the tidal locking radius, using our moon with 10 $\times$ the mass of Ganymede = 0.25 M$_{\oplus}$ and $t_L$ = 4.5 Gyr.  Comparatively, when the same test was run for a moon with M$_m$ = 10 M$_{\oplus}$, there were 44 planets with a smaller tidal locking radius.  Out of the 35 super-Ganymede-moon systems, only 3 of them spent any part of their orbit within the habitable zone: HD 4732 c (70\%), HD 73534 b (62.7\%), and HD 106270 b (20.7 \%).  The habitable zone boundaries for HD 4732 c were [4.18, 7.49] AU, which we believed 
placed the exoplanet too far from the host-star such that the heat from gravitationally-induced shrinking of the gaseous planet was no longer negligible.  Similarly,  HD 106270 b spent such a small percentage of its orbit in the habitable zone that we felt the planet and moon would be too cold.  Therefore, we have investigated HD 73534 b \citep{Valenti2009}, which is a 1.104 \MJ planet orbiting a 1.23 \MS G5IV-type star at a$_{sp}$ = 3.068 AU and e$_{sp}$ = 0.074 (Fig. \ref{hzs}, bottom).

\begin{figure*}
  \begin{center}
  \includegraphics[width=18cm]{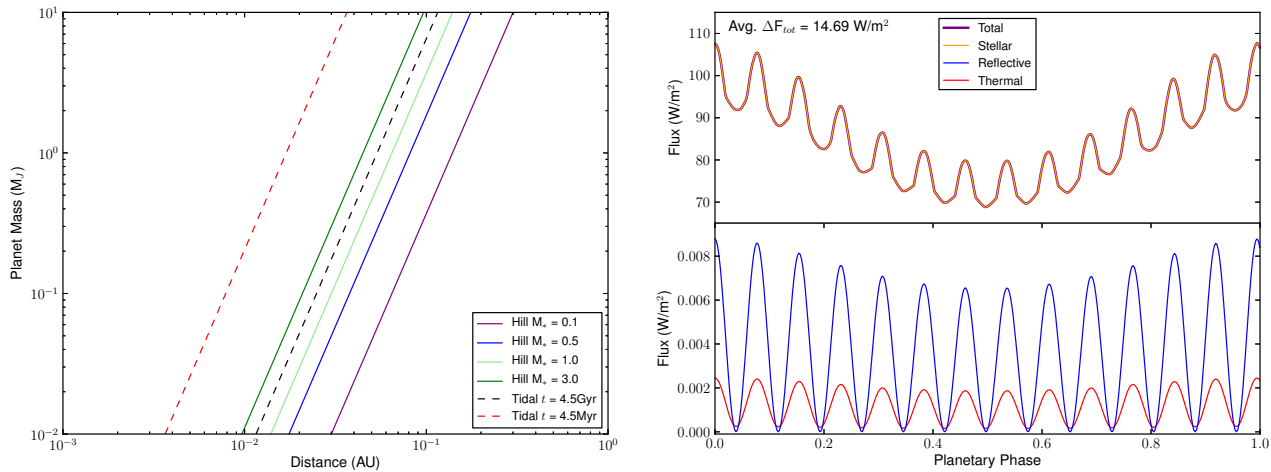}
  \end{center}
  \caption{Similar to Fig. \ref{muAra} but for HD 73534 b,  which spends $\sim$62.7\% of it's orbital phase in the habitable zone \citep{Kane2012}, where the hypothetical moon is near the tidal locking radius at a = 0.05 AU.}
  \label{hd7}
\end{figure*}

Figure \ref{hd7} (left) confirms that the tidal locking radius with $t_L$ = 4.5 Gyr is smaller than the Hill radius for a solar-type star, 0.055 AU and 0.062 AU, respectively.  Therefore, we have placed the hypothetical moon at a distance of a$_{pm}$ = 0.05 AU from the host-planet, to ensure that the moon is well within the tidal locking limits.

The flux profile for HD 73534 b is shown in Fig. \ref{hd7} (right), where $P_{pm} / P_{sp}$ = 1/13, due to the large planet-moon semi-major axis.  However, while the a$_{pm}$ for this system is the largest that we've investigated, it is tempered by the fact that a$_{sp}$ = 3.068 AU, which is also the largest planetary semi-major axis within our applications.  Therefore, we do not see the large average total flux fluctuation on the surface of the moon, $\sim$15 W/m$^2$, as seen in Fig. \ref{hd2} (right). The maximum flux experienced by the moon's surface in this system is 108 W/m$^2$, while the minimum is 69 W/m$^2$.

\section{Exomoon Habitability}
\label{hab}
We have calculated the flux phase curves for hypothetical moons based on the mean flux received on the surface.  The total irradiation may be translated into a variety of surface conditions, depending on the specifics of the moon's atmosphere, such as: composition, pressure, circulation, wind speeds, etc. \citep{Kasting:1993p6301, Selsis2007}.  
In this section we examine the potential habitability of the hypothetical exomoons in the four application systems as well as more extreme scenarios in which the planetary host may affect the temperature of the hypothetical moon.

\subsection{Four Physical Systems}
It could be said with some confidence that the incident flux gradient from the equator to the poles of the moon result in significantly lower equilibrium temperature at the pole.  This idea holds regardless of the invoked climate model, especially given that we defined the incident angle between the moon and the planet as $i$ = 0\degree for all system applications.  However, the complicated, multi-parameter climate model's dependencies make it necessary to use the incident flux as a first-order approximation for the thermal equilibrium temperature of a black body at theÊtop of the moon's atmosphere.  

We determine the habitability of the hypothetical exomoons described in \S \ref{apps} by utilizing the methodology applied by \citet{Kane2012}.  They first determined the inner and outer edges of the habitable zone based on the runaway greenhouse and maximum greenhouse effects \citep{Underwood2003}.  Then, recognizing the limited amount of information regarding the atmosphere and surface of the exoplanet (in this case exomoon), they calculated a possible range for the equilibrium temperature based on the heat redistribution of the atmosphere.  For example, if the atmosphere is assumed to be 100\% efficient, the equilibrium temperature is defined as:
\begin{equation}
T_{100\%} = \bigg( \frac{L_{\star}\,(1-\alpha_p)}{16\,\pi\,\sigma\,r^2} \bigg)^\frac{1}{4} ,
\end{equation}
where $r$ is the distance between the star and the planet-exomoon system.  In comparison, if the atmosphere is not at all efficient in redistributing the heat, such that there is a hot day-side, then $T_{0\%} = T_{100\%} \times 2^{-1/4}$.  See \citet{Kane2012} for more discussion.

In Table \ref{tab.temp} we give the thermal equilibrium temperature range corresponding to both $T_{100\%}$ and $T_{0\%}$ at the inner and outer edges of the habitable zone using \citet{Kopparapu2013} for all four of our application systems.  
 We note that the habitable zone regions for smaller-size planet and moons are slightly different than those calculated here, although not significantly.  The equilibrium temperature was calculated from the stellar flux, such that T$_{eq}$ = $(F_s/\sigma)^{1/4}$, which by far dominates the sources of flux received on the moon's surface.  

 We have included the average, minimum, and maximum flux and thermal temperatures for the hypothetical moon in each system within the table, as a comparison to the habitable zone equilibrium temperature boundaries.  The average equilibrium temperature of the moon around HD 28185 b is warmer than the inner habitable zone boundaries defined by a 100\% efficient atmospheric heat redistribution.  
 In comparison, the moons orbiting BD +14 4559 b and HD 73534 b are too cold for a 0\% efficiency rate, but have an average thermal temperature within the range for 100\% redistribution efficiency.  The hypothetical moon surrounding $\mu$ Ara b is the only moon with an average equilibrium temperature that falls within the inner and outer habitable zone boundaries for both extremes in the atmospheric heat redistribution.

The differentiation in average equilibrium temperature between the four systems is most likely due to the fractional time that HD 73534 b spends within the habitable zone which, according to \citet{Kane2012}, is $\sim$62.7\% of its orbital phase.  In addition, the moon's semi-major axis is +1.4 times the distance from the planet as compared to the other systems.  The large distances from both the star and the planet significantly reduce the radiation experienced on the moon's surface.  However, if the hypothetical moon has an atmosphere that is relatively proficient at distributing the heat, then the moon's average thermal equilibrium temperatures are habitable, even if this host-planet isn't within the habitable zone 100\% of the time.  For all the cases we've examined, the hypothetical exomoon at the tidal locking radius around HD 73534 b is the only exomoon whose equilibrium temperature range lies fully within any of the habitable zone boundaries, in this case, for a fully efficient heat redistribution.

As was discussed in \S \ref{radii}, large planetary eccentricities give rise to substantial ranges in the equilibrium temperature. The extreme thermal temperatures experienced on the moon around BD +14 4559 b, where e$_{sp}$ = 0.29, make it inhabitable no matter the heat distribution efficiency and despite the average thermal temperature.  For a fully efficient heat redistribution, the thermal temperature of the moon reaches a maximum temperature that is $\sim$8K above the inner habitable zone limit and a minimum thermal temperature $\sim$1K below the outer habitable zone limit.  In this scenario, the increased flux received by the exomoon at the time of the planet's periastron was not counterbalanced by the fact that the planet only spends $\sim$68.5\% of it's orbital phase within the habitable zone \citep{Kane2012}.  
As a result, the exomoon experiences an equilibrium temperature swing that spans the entire range of the inner and outer habitable zone limits for a fully efficient heat redistribution.  When the atmosphere is 0\% efficient, the exomoon is in general too cold to be habitable, such that the maximum temperature lies is nearly exactly halfway between the inner and outer habitable zone boundaries (see Table \ref{tab.temp}).

We find that for HD 28185, the maximum equilibrium temperatures for the hypothetical moon are above, $\sim$ 13 K, the inner habitable zone boundaries at 100\% efficiency.  
However, the thermal temperature range is fully encompassed in the inner and outer habitable zone boundaries for 0\% efficiency. 
In contrast, the moons around BD +14 4559 b and HD 73534 b have equilibrium temperatures more closely aligned with an atmosphere that is completely efficient at redistributing heat.  The thermal temperature ranges experienced by both moons are almost fully within the inner and outer habitable zone boundary at 100\% heat redistribution.  It is only the exomoon orbiting around $\mu$ Ara b that is too hot when there is fully efficient heat redistribution (maximum thermal temperature 5 K above the inner habitable zone) and also too cold when there is no heat redistribution (minimum thermal temperature 9 K below the outer habitable zone).  In other words, the thermal equilibrium temperatures would fall within the inner and outer habitable zone boundary for a heat redistribution between 0--100\%.

We have analyzed four physical systems with hypothetical moons near the Hill or tidal locking radius.  By examining the thermal equilibrium temperature range imposed by the geometry of the system, namely the minimum, maximum, and average, we find that the equilibrium temperature of the moons fall within habitable limitations during two scenarios.  The first scenario occurs when the host-planet is fully within the habitable zone and the heat redistribution on the moon is relatively inefficient ($<$ 50\%).  Second, the moon is habitable when the host-planet spends only a fraction of its phase in the habitable zone and the heat redistribution on the moon is closer to 100\% efficiency.  In those cases where the equilibrium temperature extrema brought on by high planetary eccentricity cannot be subdued by less time (or orbital phase) in the habitable zone, the exomoons may not be habitable.  By looking specifically at exomoons at the furthest stable/locked radii from the host-planet, we have placed an upper bound on the range of expected thermal equilibrium temperature profiles for rocky moons.

\subsection{Extreme Flux Contributions}
The thermal equilibrium temperature of an exomoon is affected by the both the host-star and -planet.  However, there are a number of extreme scenarios that would maximize the received flux on the moon and consequently increase its equilibrium temperature towards habitability.  If we were to change the parameters of the star-planet-moon system, then an increased stellar luminosity or planetary radius would bolster the thermal temperature of the exomoon.  Decreasing the distance between the planet-moon system and the star, or even the distance between the planet and the moon, would also have a similar effect.  In addition, if we allowed the exomoon's eccentricity to become non-zero, then there would be an increase in tidal heating between the planet and the moon \citep{Heller2013}.

However, given the confines of our study which has been to analyze physical, stable star-planet systems where the hypothetical moons are located relatively far from the host-planet, then we are left with few ways in which to maximize the equilibrium temperature of the exomoon from contributions by the host-planet.  If the planet was relatively young and enriched with radioactive isotopes (such as $^{26}$Al or $^{60}$Fe), then the temperature of the planet would be greatly elevated to the point where volatiles were evaporated from the surface \citep{Grimm:1993p4103}.  The exomoon would experience this thermal planetary increase to the power of four via the flux, however, the short-lived nature of the isotopes would mean that the planet and exomoon would cool after $\sim$1 Myr.  A giant impact colliding with the planet may also have a similar effect, i.e. causing the planet's temperature to drastically rise, perhaps through the release of magma, but then eventually cool with time.  Another scenario depends on whether the planet is tidally locked to the star, creating a distinct bright and dark side.  The thermal flux received by the exomoon would be increased if the orbital angle of the exomoon was adjusted in such a way as to maximize the contribution of the bright side's significantly warmer (as opposed to the dark side's) thermal irradiation onto the exomoon.  Finally, changing the albedo of planet would alter both the thermal and reflected planetary flux contribution onto the exomoon, for example by a large impact or on a terrestrial planet via volcanic eruptions or vegetation.  Through these more extreme scenarios, the host planet alone may raise the equilibrium temperature of the exomoon towards habitability, although either for relatively short periods of time or through less significant contributions.

\section{Future Detections}
\label{detect}
Detection of exomoons is one of the next thresholds to be traversed
in exoplanetary science. However, the presence of an exomoon within a 
planetary system will produce a negligible effect in most current detection 
techniques. For all four applications described in \S \ref{apps}, the center-of-mass of the planet-moon
system lies inside the radius of the planet itself. Thus, detection
via radial velocity or astrometry effects is impractical due to its
negligible effect on the host star.  
Microlensing has also been proposed by \citet{Han2008} as a method to search for exomoons. However, this technique is most sensitive to moon-planet distances beyond 0.05 AU for a Jupiter-mass planet, which is outside the separation range we consider. Additionally, the lack of follow-up opportunities presented by microlensing discoveries makes characterization of those moons difficult.

Comparatively, the transit technique is one method which may present opportunities to
search for detectable exomoon signatures in the near future.  Attempts to use transits are
already underway using the Kepler data \citep{Kipping2012, Kipping2013}, although
these searches are restricted to the relatively short orbital periods
where the detections are more frequent. We investigated the potential
transit parameters for the \S \ref{apps} applications by calculating the
expected transit probability, duration, and depth for an edge-on 
planetary orbital orientation, shown in Table \ref{transit}.

There are many possible locations for the moon, with respect to planet, during a transit.
Each location has its own distinct signature, some of which blend with the
transit of the host planet \citep{Kipping2011}.  We have examined the epoch 
when the planet and moon are at a maximum angular separation, such that the moon
either leads or trails the planet, when calculating the separation between the transits, also shown 
in Table \ref{transit}.  In most cases, there will be times when the planet
and the moon have completely separate transits across the face of the
host star. The exception to this is BD +14 4559 b, where the smaller
orbital period of the planet results in an transit overlap between the planet and moon,
though they are at maximum angular separation.

Future prospects for detection of exomoon transits are exceedingly
difficult and will require photometric precision better than
$10^{-5}$. As shown in
Table \ref{transit}, the predicted transit depths of the moons push heavily
against the boundaries of what is achievable with current ground and
space-based instruments. Some of the best cases for high precision studies will be instances
of planets discussed in this paper: giant planets
orbiting bright host stars, which would present unique opportunities if found to transit.
Improving the measured orbits of long-period planets specifically in the habitable zone is
already being undertaken by a number of projects, such as the Transit Ephemeris
Refinement and Monitoring Survey (TERMS), with an aim of detecting
transits of known exoplanets \citep{Kane2009, Wang2012}. A concerted effort to search the moons around
HZ giant planets will therefore likely need to await future generation
telescopes, such as the European Extremely Large Telescope,
the Thirty Meter Telescope, the Giant Magellan Telescope,
and the James Webb Space Telescope.

\section{Summary}
\label{sum}
In order to better understand the potential habitability of exomoons, we have simulated the flux phase curves of a hypothetical exomoon orbiting an exoplanet for the duration of one planetary orbit.  Our calculations employed the technique used by \citet{Heller2013} to best estimate the heat contribution by the host-star as well as the host-planet, where the moon receives radiation from the planet that is both thermal and reflected.  Given that eccentricity tends to increase with planetary mass, as has been noted in the sample of confirmed RV exoplanets, we analyzed the effect of a large eccentricity on the moon's flux phase profile.  For relatively large eccentricities (e$_{sp} >$ 0.2), the changing distance between the host-planet and star dominates the orbit of the moon around the planet and significantly varies the equilibrium range on the moon.  Finally, we examined the limit between the Hill and tidal locking radii, since the proximity of the exomoon to the planet has a large impact on the radiation seen from the star and planet.

The point of this study is to quantitatively show the contributions to the flux on an exomoon in a system with a massive planet, which will be common, but has yet to be thoroughly examined.  While we have explored systems with moons at their largest stable distance from their host planet, let us assess the other extrema with respect to our Example system.  The Roche limit in the Example (see \S \ref{illrad}) is roughly 0.001 AU for a rigid satellite/moon.  As per the rest of our paper in order to err on the side of caution, this places the moon at a$_{pm}$ = 0.006 AU.  However, the smallest stable distance for the moon is less than two times the distance that we originally used in our example, where a$_{pm}$ = 0.01 AU.  Given that our Example system had a planet that was on the lower end of the masses and therefore a smaller Roche limit, the planetary contribution to the exomoon will always be relatively low when analyzing the averages, as we have done here.  However, that does not mean that the planetary contributions are negligible to the overall flux experienced at the top of the moon's atmosphere -- see \citet{Heller2013}, their Figs. 7, 11, and 12.  Instead, our model takes into account the extrema experienced on the exomoon, over all latitudes and longitudes and throughout the orbit of the moon about the star.  While we average them in order to determine the average flux, the minimums and maximums experienced on the exomoon affect the overall average flux which result in more accurate simulations.

We applied our analysis to four physical systems: $\mu$ Ara b, HD 28185 b, BD +14 4559 b, and HD 73534 b, where we simulated the same moon in each system at it's Hill or tidal locking radius, whichever was more limiting.  Despite the larger, brighter host star for $\mu$ Ara b and smaller semi-major axis between the planet and the moon, we found that the average equilibrium for the exomoon around HD 28185 b was higher due to the closer proximity of the star to the planet.  These two exomoons, plus the moon around BD +14 4559 b, experienced fluxes which were noticeably higher with respect to the moon around HD 73534 b.  We find the fractional orbital phase ($\sim$62.7\%) that HD 73534 b spent in the habitable zone, coupled with the large semi-major axis between the moon and the planet, led to much lower fluxes on the moon.  

We analyzed the habitability of the hypothetical moons around the four application systems by invoking a first-order approximation for the thermal temperature 
using the stellar flux only.  
In lieu of a multi-parameter climate model, we determined a range of temperatures for each exomoon based upon the heat redistribution efficiency of the surface, varying from 0-100\% efficiency. 
 The average temperature for the HD 28185 b exomoon with 100\% efficiency fell outside the inner habitable zone.  The temperature ranges of the moons around $\mu$ Ara b and HD 28185 b were better suited for a less efficient heat redistribution, although the minimum temperature still fluctuated below the outer habitable zone temperature in the case of $\mu$ Ara.  The temperatures experienced on the BD +14 4559 b exomoon were far too cold for 0\% heat efficiency, but were better encompassed by a 100\% heat efficiency.  Out of all four systems, the exomoon around HD 73534 b was the only instance where the entire temperature range was within the habitable zone boundary for a fully efficient heat redistribution.  Therefore, we conclude that exomoons that traverse outside of the habitable zone require a redistribution efficiency close to 100\% in order to be habitable.  
 In addition, if the host-planet were to undergo more extreme conditions, such as radiogenic heating, giant impact, or a change in its albedo, or if the angle of the moon-orbit was preferential to the bright-side of a tidally locked planet, then the exomoon may achieve temperatures more favorable to habitability.

Because of the relatively small size of the exomoon in comparison to the host-star and -planet, the majority of exoplanetary detection techniques are not applicable.  Observing transiting exoplanets may be the most optimistic method for discovering an exomoon, to-date.  However, in order to be successful, this method relies upon a specific geometry between the planet and moon during transit (maximum angular separation) as well as extremely high photometric precision.  Fortunately, systems similar to the applications we've explored, with giant planets orbiting bright stars, may be the the most likely candidates for detecting an exomoon.

\section*{Acknowledgements}

The authors acknowledge financial support from the National Science
Foundation through grant AST-1109662.  This work has made use of the
 Habitable Zone Gallery at hzgallery.org.  The authors would like to thank
Ren\'{e} Heller for use of his publicly accessible and well documented
code, {\it exomoon.py}, as well as his useful
suggestions and insight.  NRH thanks CJ Messinger, Esq., for his
inspiration and CHW3 for his support.

\begin{deluxetable*}{c|ccccccc}
Ê \tablecolumns{8}
Ê \tablewidth{0pc}
Ê \tablecaption{\label{tab.params} Parameters for the Exoplanet-Moon Systems}
Ê \startdata
\hline
\hline
System & Stellar Mass & Stellar Radius & T$_{eff}$ &  Planet Mass & a$_{sp}$ & e$_{sp}$ & a$_{pm}$ \\ 
Name  & $(M_{\odot})$& ($R_{\odot})$   & $(K)$        & $(M_{Jup})$   & $(AU)$      &               & $(AU)$ \\
\hline
Example &  1.0 & 1.0 & 5778 & 1.0 & 1.0 & 0.0/0.4 & 0.01/0.02 \\

$\mu$ Ara b & 1.15  &  1.25  & 5784   & 1.746 & 1.527  & 0.128  & 0.03 \\

HD 28185 b & 0.99 & 1.007 &  5656 & 5.80 & 1.023 & 0.05 & 0.035 \\

BD +14 4559 b & 0.86 & 0.726 & 4814 & 1.52 & 0.776 & 0.29 & 0.01 \\

HD 73534 b &  1.229 & 2.288 & 5041 & 1.104 & 3.068 & 0.074 & 0.05 \\
\hline
\hline
\enddata
\tablenotetext{$^*$}{ For all systems: R$_p$ = 1.0 R$_{Jup}$, $\alpha_p$ = 0.343, M$_m$ = 10 $\times$ the mass of Ganymede = 0.25 M$_{\oplus}$, R$_m$ = 0.68 R$_{\oplus}$, moon rock-to-mass fraction = 0.68, e$_{pm}$ = 0.0, $i$ = 0\degree, and $\omega$ = 0\degree.  For the Example system, M$_m$ = M$_{\oplus}$ and R$_m$ = 0.09 R$_{\oplus}$.}
\end{deluxetable*}

\begin{deluxetable*}{l|ccc|cccccc}
Ê \tablecolumns{10}
Ê \tablewidth{0pc}
Ê \tablecaption{\label{tab.temp} Exomoon Flux (Total) and Equilibrium Temperature (Stellar Only)}
  \tablehead{
    \colhead{} &
    \multicolumn{3}{c}{Flux  $(W/m^2)$ } &
    \multicolumn{6}{c}{T$_{eq}$ from Stellar Flux $(K)$} \\
    \colhead{Name} &
    \colhead{Avg} &
    \colhead{Min}  &
    \colhead{Max} &
    \colhead{Avg} &
    \colhead{Min} &
    \colhead{Max} &
     \colhead{Location} &
    \colhead{T$_{100\%}$} &
    \colhead{T$_{0\%}$} \\
   }
Ê \startdata
$\mu$ Ara b & 176 & 130 & 256 &  236 & 219 & 259 &Inner HZ & 254 & 302\\
              &  & & & & & & Outer HZ & 192  & 228 \\
HD 28185 b & 234 & 194  & 295 & 253 & 242 & 268 & Inner HZ & 253 & 301 \\
&  & & & & & & Outer HZ & 191 & 227 \\
BD +14 4559 b & 113 & 63  & 241 & Ê209 & 183 & 255 & Inner HZ Ê& 247 &  294 \\
 &   & &    &Ê& & & Outer HZ Ê& 184  & 219 \\
HD 73534 b & 84 & 69 & 108  & 196 & 187 & 208 & Inner HZ &Ê 249 &  296 \\
Ê & & &  &  & & & Outer HZ &Ê186 &  221 \\
\hline
\hline
\enddata
\end{deluxetable*}

\begin{deluxetable*}{l|cccccc}
  \tablecolumns{7}
  \tablewidth{0pc}
  \tablecaption{\label{transit} Potential System Transit Parameters}
  \tablehead{
    \colhead{} &
    \colhead{} &
    \multicolumn{2}{c}{Planet} &
    \multicolumn{2}{c}{Moon} &
    \colhead{}  \\
    \colhead{System} &
    \colhead{Prob} &
    \colhead{Duration}  &
    \colhead{Depth} &
    \colhead{Duration} &
    \colhead{Depth} &
    \colhead{Separation} \\
     \colhead{} &
    \colhead{$(\%)$} &
    \colhead{$(days)$}  &
    \colhead{$(\%)$} &
    \colhead{$(days)$} &
    \colhead{$(\%)$} &
    \colhead{$(days)$}
   }
  \startdata
$\mu$ Ara b	&	0.43  &	0.771   &	  0.728  &  0.714	&   0.003 &   2.011 \\
HD 28185	 b  &  0.51&	0.569&	  1.115&  0.518&	   0.004 & 2.064 \\
BD +14 4559  b &	0.68	&0.303&	  2.193&  0.266	 &  0.008&  0.552 \\
HD 73534	 b  &0.36	&1.935	&  0.217  & 1.864	&   0.001 & 4.591 \\
\hline
\hline
 \enddata
\end{deluxetable*}

\end{document}